\documentclass[aps,pra,10pt,showpacs,amsmath,amssymb,floatfix,superscriptaddress,twocolumn]{revtex4-2}
\usepackage{physics,graphicx,microtype}
\usepackage[dvipsnames,svgnames,table]{xcolor}
\usepackage{tcolorbox}
\usepackage[unicode=true,
            pdfusetitle,
            bookmarks=false,
            breaklinks=true,
            pdfborder={0 0 0},
            backref=false,
            colorlinks=true,
            hypertexnames=false]{hyperref}
\hypersetup{linkcolor=NavyBlue,urlcolor=NavyBlue,citecolor=NavyBlue}

\usepackage{mathrsfs,esint,dsfont}
\usepackage{newtxtext,newtxmath,fix-cm}
\DeclareMathAlphabet{\mathcal}{OMS}{cmsy}{m}{n}
\newcommand{\devnull}[3]{}

\begin{document}

\title{Measuring the Oscillation Frequency Beyond the Diffraction Limit}
\author{Chao-Ning Hu}
\author{Jun Xin}
\email{jxin@hdu.edu.cn}
\author{Xiao-Ming Lu}
\email{lxm@hdu.edu.cn}
\affiliation{School of Sciences and Zhejiang Key Laboratory of Quantum State Control and Optical Field Manipulation, Hangzhou Dianzi University, Hangzhou 310018, China}

\begin{abstract} 
High-resolution array detectors are widely used in single-particle tracking, but their performance is limited by excess noise from background light and dark current.
As pixel resolution increases, the diminished signal per pixel exacerbates susceptibility to noise, degrading tracking accuracy.
To overcome this limitation, we use spatial-mode demultiplexing (SPADE) as a noise-robust approach for estimating the motion characteristics of an optical point-like source.
We show that SPADE efficiently concentrate the information into a few key spatial modes, drastically reducing the number of detectors while maintaining high estimation precision. 
Furthermore, we enhance the robustness of the estimation against excess noise by elaborately designing the modes to be decomposed. 
We demonstrate, both theoretically and experimentally, that a SPADE with two specific modes outperforms direct imaging in estimating the micro-oscillation frequency of an optical point source in the presence of excess noise.
\end{abstract}
\maketitle

\textit{Introduction.---}
Single-particle tracking (SPT) is a crucial technique in modern time-resolved microscopy, enabling precise observation of nano-scale motion or fluctuation of light-emitting sources, such as single molecules, viruses, or quantum dots~\cite{Levi2007,Mortensen2010,Deschout2014,Shen2017}.
Conventional SPT begins by capturing a sequence of time-resolved images of a single point-like source through high spatial-resolution direct imaging.
The center of the particle is then identified in each image, which facilitates the determination of its trajectory, vibrational frequency, and real-time velocity~\cite{Cheezum2001,Carter2005}.
Among the key parameters extracted from SPT, the oscillation frequency is a critical descriptor of dynamic behavior.
In biological systems, the oscillation frequencies of particles such as viruses and bacteria reflect important aspects of their mechanical and physiological states~\cite{Zhang2025}.
In astrophysics, periodic deviations in the trajectories of celestial bodies---known as perturbations---result from gravitational interactions with nearby masses and are essential for understanding orbital mechanics, predicting celestial events, and detecting unseen objects such as exoplanets~\cite{Mouret2007}.
The reliability of these analyses, however, is largely dependent on the localization accuracy of the single optical point source in each frame within the exposure time.
Such localization accuracy is limited by various factors in practical implementation, including diffraction of light, photon shot noise, photon collection efficiency, and other experimental imperfections.

The localization of optical point source is significantly influenced by diffraction effects, which causes even infinitely small point sources to appear as finite-sized spots in the image plane.
While diffraction traditionally sets a limit on resolution of direct imaging~\cite{LordRayleigh1879,Ram2006}, experimental imperfections often play a more substantial role in determining the localization accuracy than theoretical limits~\cite{Deschout2014,Thompson2002,Yildiz2003}.
Factors such as the finite pixel size of the detector array and excess noise from background light and dark current can significantly degrade the localization precision.
Large pixels reduce precision by obscuring the exact position of photons within each pixel, while small pixels increase the likelihood of signal being overwhelmed by noise, making the localization process more vulnerable to excess noise.
To overcome these limitations, a promising strategy involves concentrating the point source's relevant information onto a minimal number of detector pixels. 
This approach necessitates the development and implementation of sophisticated measurement schemes.

One of the most promising measurement schemes is spatial-mode demultiplexing (SPADE), proposed by Tsang \textit{et al}.~\cite{Tsang2016,Tsang2020} in 2016 to address the longstanding challenge of resolving two incoherent optical point sources in the sub-Rayleigh regime.
Unlike direct imaging, which measures the positions of individual photons, the SPADE measurement separates the image-plane optical field into nontrivial spatial modes, such as Hermite-Gaussian (HG) modes.
This approach has been shown to significantly improve the precision of estimating the separation between two incoherent point sources and reduce the error probability of testing hypotheses about the properties of optical sources~\cite{Lu2018,Huang2021,Bao2021}.
Experimentally, SPADE has been implemented using various techniques, e.g., inversion of coherence along an edge~\cite{Tham2017,Wadood2024}, heterodyne detection~\cite{Yang2016,Pushkina2021,Frank2023,Duplinskiy2024_published}, digital holography~\cite{Paur2016,Zhou2023,Hu2023,Peterek2023}, multi-plane light conversion~\cite{Boucher2020,Tan2023,Santamaria2023,Rouviere2024,Santamaria2024,Ozer2024}, and nonlinear optics~\cite{Darji2024}. 
While SPADE has been primarily applied to resolve static characteristics of optical point sources, its capability for measuring motion characteristics such as temporal frequency under noisy conditions remains largely unexplored.

In this work, we use SPADE for estimating the parameters characterizing the motion of a single optical point source.
Through quantum parameter estimation theory, we demonstrate that SPADE enables robustness against excess noise in the estimation of optical source motion characteristics while maintaining high precision, outperforming conventional imaging in this crucial aspect.
Our analysis reveals that in the sub-Rayleigh regime, the essential motion information can be predominantly captured by a few key spatial modes to be measured, thereby suppressing the total amount of excess noise. 
Our experiment focuses on the challenge of estimating the oscillation frequency of an optical point source in the sub-Rayleigh regime with noise---a task central to the SPT application.
By strategically designing the spatial modes for decomposing the image-plane optical field, we demonstrate that SPADE exhibits superior robustness against excess noise in the micro-oscillation frequency estimation when compared to direct imaging.

\begin{figure}[t]
  \centering
  \includegraphics{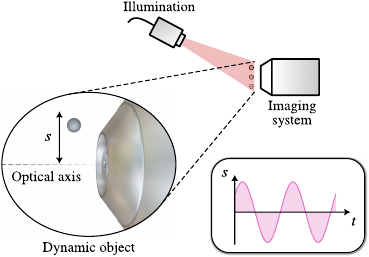}
  \caption{Illustration of the dynamic single point source.}
  \label{fig:dynamic}
\end{figure}

\textit{Theoretical analysis.---}
Let us consider an optical point source, or an illuminated particle, transversely moving away from the optical axis in the object plane, as illustrated in Fig.~\ref{fig:dynamic}.
For simplicity, we assume that the motion of the point source is one-dimensional.
Let us denote the displacement of the point source from the optical axis by a function $s(t, \theta)$, where $t$ denotes time and $\theta$ represents the unknown parameters that characterize the motion, such as the frequency, amplitude, and initial phase in the case of simple harmonic motion.
Our goal is to estimate the unknown parameters $\theta$ as precisely as possible by observing the optical fields on the image plane.

We use quantum parameter estimation theory to analyze the performance of a measurement strategy for estimating the motion characteristics \( \theta \).
The covariance matrix \(\operatorname{Cov}(\hat\theta)\) of any unbiased estimator for a vector parameter \( \theta \) must obey the inequalities \( \operatorname{Cov}(\hat\theta) \geq F^{-1} \geq \mathcal{F}^{-1} \), where \( F \) and \( \mathcal{F} \) are the classical Fisher information (CFI) matrix and the quantum Fisher information (QFI) matrix, respectively~\cite{Helstrom1967,Helstrom1968,Helstrom1976,Liu2020,Paris2009}.
Here, the matrix inequality is interpreted as the Loewner ordering of positive semidefinite matrices~\cite{Horn2012,Bhatia1997}, that is, \( A\geq B \) means that \( A-B \) is positive semidefinite.
In the case of single-parameter estimation, the CFI quantifies the amount of information about the unknown parameter that a specific measurement provides, while the QFI represents the maximum CFI achievable over all possible measurements.

We assume that the point source is a weak thermal optical emitter, such that the density operator for the optical fields on the image plane in each short coherence time interval can be expressed as \( \rho_t \approx (1-\epsilon) \op{\mathrm{vac}} + \epsilon \op{\psi_t} \), where \( \epsilon \ll 1 \) is the average photon number per coherence time interval, \( \ket{\mathrm{vac}} \) denotes the vacuum state, and \( \ket{\psi_t} \) is the one-photon state.
Here, \(\epsilon\) is assumed to be time-independent.
We consider a scalar optical field at the image plane of a diffraction-limited imaging system~\cite[Sec.~6.1]{Goodman2003}. 
The performance of a diffraction-limited imaging system, e.g., high-end telescopes and microscopes, is limited only by the diffraction of light, without any additional aberrations or imperfections, so that we can focus the optimization on detection strategies implemented at the image plane.
For theoretical simplicity, we assume that the imaging system is one-dimensional and spatially invariant~\cite[Sec.~2.3]{Goodman2003} so that the normalized point-spread function (PSF) is independent of the position of the point source in the object plane and denoted by \(\psi(x)\) with \(x\) being the transverse coordinate on the image plane.
In this case, when the point source undergoes a displacement \( s \), the one-photon state can be expressed as \( \ket{\psi_t} = \int \psi(x-s(t,\theta)) \ket{x} \dd{x}\), where \(\ket{x}\) is the photon image-plane position eigenket.

To estimate the parameters associated with the motion of the light source, measurements must be taken at different time instants.
Let \(T\) denote the set of sampling time instants.
We assume that the motion of the point source is sufficiently slow such that its position remains approximately constant during each sampling process, such as the exposure time of a camera.
Under this assumption, the total density operator for all the samples can be expressed as \( \rho_\mathrm{total} = \bigotimes_{t\in T} \qty(\rho_t^{\otimes M}) \), where \(M\) is the number of temporal modes within a single sampling process.
Furthermore, we assume that the PSF can be approximated by a Gaussian function $\psi(x) = \qty(2\pi\sigma^{2})^{-1/4}\exp(-x^{2}/(4\sigma^{2}))$, where $\sigma$ denotes the characteristic width of the PSF. 
Based on this assumption, we get the QFI matrix about the vector parameter \( \theta \)~\cite{SupplementalMaterial}\nocite{Wasserman2010, Stoica1989, Kay1993, RosalesGuzman2017, Arrizon2007}:
\begin{equation}\label{QFIM}
  \mathcal{F}_{jk} \approx \frac{\nu}{\sigma^2} \sum_{t\in T} \pdv{s(t,\theta)}{\theta_j} \pdv{s(t, \theta)}{\theta_k}
\end{equation}
with \( \nu = \epsilon M \) being the average photon number collected within each sampling.

A concrete quantum measurement at each temporal mode can be described by a positive-operator-valued measure (POVM) \( \{E_j\} \) defined on the Hilbert space associated with one-photon states.
Each photon detectors corresponds to an element of the POVM, representing the event that a photon is detected.
At each sampling instance, the multi-output photon counting measurement yields a measurement outcome \(m=(m_1, m_2, \ldots)\), where \(m_j\) denotes the integrated photon number recorded by the \(j\)th photon detector.
It is standard to assume that each \(m_j\) follows a Poisson distributions~\cite{Tsang2021, Smith2010, Starr2012}.
Additionally, we account for the background noise of the photon counters, which may arise from sources such as background light or dark current.
Since the photon count due to background noise in each temporal mode is a rare event, the background noise in the \( j \)th detector during a single sampling process can also be assumed to follow a Poisson distribution with an average photon number \(b_j\).
Under this assumption, we get the CFI matrix~\cite{SupplementalMaterial} as
\begin{equation}\label{eq:CFI}
  F_{jk}
  = \nu \sum_{t \in T} \gamma(t,\theta,b) \pdv{s(t,\theta)}{\theta_j} \pdv{s(t,\theta)}{\theta_k},
\end{equation}
where
\begin{equation} \label{eq:gamma}
    \gamma(t, \theta, b) = \sum_j \frac1{\mu_j + b_j / \nu} \qty(\pdv{\mu_j}{s})^2
\end{equation}  
with \( \mu_j = \ev{E_j}{\psi_t} \).
The CFI matrices for any measurement under the noisy conditions are bounded from below as \(F \leq \mathcal F\), where the QFI matrix under the same noisy conditions is given by~\cite{SupplementalMaterial}
\begin{equation} \label{eq:noisy_qfi}
    \mathcal{F}_{jk} = \frac{\nu}{\sigma^2}\frac{1}{1+2b/\nu}\sum_{t\in T} \pdv{s(t,\theta)}{\theta_j} \pdv{s(t,\theta)}{\theta_k}.
\end{equation}

The quantity \( \gamma(t,\theta,b) \) defined in Eq.~\eqref{eq:gamma} plays a crucial role in analyzing both the optimality and robustness of a quantum measurement for estimating the parameters of a dynamic point source.
In the absence of background noise, i.e., when \( b_j = 0 \) for all \( j \), \( \gamma(t,\theta,b) \) reduces to the CFI per detected photon with respect to the displacement \( s \) of the point source.
As shown in Ref.~\cite{Zhou2023}, any measurement scheme based on a complete set of real-valued mode functions is optimal for estimating the displacement of the point source.
However, the robustness of these measurement schemes varies significantly when background noise is introduced.

Specifically, we investigate three schemes of observing the optical field on the image plane: direct imaging, Hermite-Gaussian-mode SPADE (HG-SPADE), and plus-minus-mode SPADE (PM-SPADE).
The HG-SPADE measurement counts the photons of the optical field in Hermite-Gaussian modes \(\ket{\psi_q}\) with the wave function
\begin{equation}
  \phi_q(x) = \qty(\frac1{2\pi\sigma^2})^{1/4}\frac1{\sqrt{2^qq!}}H_q\qty(\frac x{\sqrt2\sigma})e^{-x^2/(4\sigma^2)},
\end{equation}
where $H_q$ denotes the Hermite polynomial. 
The PM-SPADE counts the photons in the PM modes \( \ket{\phi_\pm} \equiv (\ket{\phi_0} \pm \ket{\phi_1}) / \sqrt2\) and discards the photons in all other modes that are orthogonal to \(\ket{\phi_\pm}\).
We denote by \( F^\mathrm{(DI)} \), \( F^\mathrm{(HG)} \), and \( F^\mathrm{(PM)} \) the CFI matrices for direct imaging, HG-SPADE, and PM-SPADE, respectively.
In the absence of background noise and for Gaussian PSFs, we have~\cite{SupplementalMaterial}:
\begin{equation}
  \mathcal{F} = F^\mathrm{(DI)} = F^\mathrm{(HG)} \geq F^\mathrm{(PM)}.
\end{equation}
Although PM-SPADE is not optimal for estimating the motion of a point source, it requires only two photon detectors and performs near-optimally when the displacement of the point source is in the sub-Rayleigh regime.
Moreover, the PM-SPADE has the same CFI matrix as the HG-SPADE with only the lowest two modes.

The robustness of a measurement scheme against background noise can be analyzed by utilizing the quantity \( \gamma(t,\theta,b) \) as follows.  
Assume that all the photon counters are subject to the same background noise, i.e., \( b_j = b \) for all \( j \).
In the case of direct imaging, the CFI is distributed across a wide range of pixels, each of which is affected by background noise.
For the HG-SPADE, the CFI is concentrated on the \(\phi_1\) mode but is very vulnerable to background noise.
This vulnerability arises because the factor \(1/(\mu_1+b/\nu)\) in Eq.~\eqref{eq:gamma} abruptly decreases when \(b\) increases from zero, as \(\mu_1\) is close to zero for small displacements.
In contrast, the PM-SPADE has two advantages to reduce the impact of background noise.
First, the PM-SPADE uses only two photon detectors, which significantly reduces the total amount of background noise photons.
Second, for small displacements, \(\mu_\pm \to 1/2\) when \(s\) approaches zero, which results in a relatively small decrease in the factor \(1/(\mu_\pm+b/\nu)\) when \(b\) increases from zero.
When the amplitude of the oscillation is very small (i.e., \(s(t, \theta) \ll \sigma\) for all \(t\)), the PM-SPADE is  optimal in the sense that its CFI attains the QFI given by Eq.~\eqref{eq:noisy_qfi}.

We emphasize that the CFI matrix and the QFI matrix given by Eqs.~(\ref{QFIM}--\ref{eq:noisy_qfi}) are valid for any deterministic parameters characterizing the motion.
Therefore, the above analysis on the robustness of different measurement schemes against excess noise applies to estimating general motion characteristics of an optical point source.

\begin{figure}[b]
  \centering
  \includegraphics{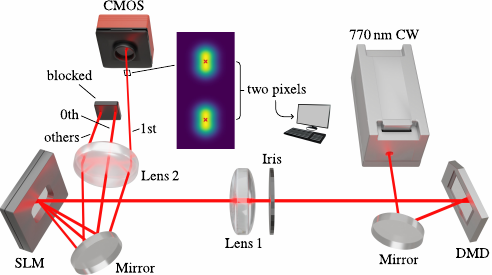}
  \caption{Experimental setup for PM-SPADE.
  The notations used are as follows:
  SLM, spatial light modulator;
  DMD, digital micromirror device;
  CMOS, complementary metal oxide semiconductor;
  CW, continuous wave.
  In this configuration, only the first-order diffracted light from the SLM is directed towards the CMOS camera. 
  The estimation process relies on intensity measurements from two specific pixels, marked with red crosses in the captured image.
  }\label{fig:setup}
\end{figure}

\textit{Experimental setup.---}
Our experiment focuses on estimating the oscillation frequency of an optical point source in the sub-Rayleigh regime with noise.
The optical point source is simulated by illuminating a light beam onto a digital micromirror device (DMD) with only one micromirror flipped.
By controlling the flipping of different micromirrors, we simulate the square-wave oscillation of an optical point source, described by $s=A+A \operatorname{sgn}[\sin(2\pi f_o t)]$, where $A$ and $f_{o}$ represent the amplitude and the frequency of the oscillation, respectively.
Note that we choose the minimum position of the oscillation as the origin of the coordinate system for convenience. 
The light emitted from the DMD is directed sequentially through an iris and a lens, constituting a unit-magnification diffraction-limited imaging system, whose PSF can be approximated to be Gaussian with a characteristic width of $\sigma \approx 103$\,{\textmu}m.

We implemented the PM-SPADE measurement using a digital holographic technique~\cite{Paur2016,Zhou2023}.
A phase-only spatial light modulator was positioned in the image plane of the imaging system and programmed to display a computer-generated hologram specifically designed for the PM-SPADE.
A CMOS camera with a sampling rate of $f_s=20$\,Hz was placed at the Fourier plane of Lens\,2 to count the photons at two specific pixels, indicated by red crosses in the inset of Fig.~\ref{fig:setup}, enabling real-time monitoring of photon occupancy in the PM modes.
We utilized 50 frames, sampled over a total duration of 2.5\,s, to perform a frequency estimation.
For each frame, we used the maximum likelihood estimator (MLE) to estimate the displacement of the point source.
Subsequently, the least squares estimation (LSE) was used to infer the oscillation frequency of the point source.
We leave the details of the experiment in Ref.~\cite{SupplementalMaterial}

For comparison, we conducted direct imaging by directly placing a CMOS camera in the image plane.
The exposure time, sampling rate, and total sampling duration were kept identical to those used in the PM-SPADE measurement.
Similar to the PM-SPADE measurement, the MLE was applied to estimate the displacement of the point source for each frame, followed by the LSE to determine the oscillation frequency of the point source.

To compare the robustness of PM-SPADE and direct imaging against background noise, we placed a brightness-adjustable LED light in front of the camera to introduce excess noise for both two schemes.
The average photon number of the background light was controlled by adjusting the LED's brightness and was experimentally determined based on the photon count recorded by the camera in the absence of the point source.

\begin{figure}[b]
  \centering
  \includegraphics{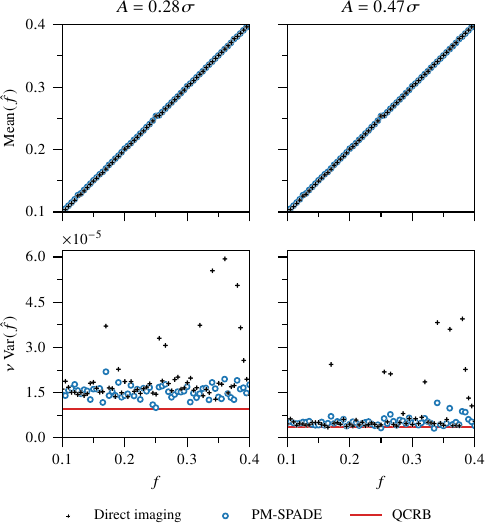}
  \caption{
    Means and rescaled variances of the PM-SPADE and direct imaging for frequency estimation in the absence of background noise.
    Here, the variance is multiplied by the average photon number \(\nu\) per frame.
    The left two panels and the right two panels correspond to the oscillation amplitude of $0.28\sigma$ and $0.47\sigma$, respectively.
    The QCRB is plotted according to Eq.~\eqref{eq:QCRB} with \(N=50\).
  }
  \label{fig:ideal}
\end{figure}

\textit{Experimental results.---}
Figure~\ref{fig:ideal} plots the means and the rescaled variances of the frequency estimates with the PM-SPADE and direct imaging in the absence of background noise.
For convenience, we henceforth use the dimensionless frequency $f:=f_{o}/f_{s}$ as the parameter of interest, where $f_{o}$ is the oscillation frequency of the point source and $f_{s}$ is the sampling rate of the CMOS camera.
The PM-SPADE and direct imaging are both repeated 200 times to evaluate the mean and variance of the estimates.
As shown in the upper two panels of Fig.~\ref{fig:ideal}, the mean values of the frequency estimates obtained by the PM-SPADE and direct imaging both match the true values of the dimensionless frequency very well.
To compare our experimental variance with its quantum limit, we calculate an approximate version of quantum Cram\'{e}r-Rao bound (QCRB)~\cite{SupplementalMaterial}:
\begin{equation} \label{eq:QCRB}
  \nu \operatorname{Var}(\hat f) \gtrsim \frac{3\sigma^2}{16 A^2 N(N-1)(2N-1)},
\end{equation}
where \(N\) is the number of samples used in each run of estimation.
As shown in Fig.~\ref{fig:ideal}, both variances of the PM-SPADE and direct imaging approach the QCRB.
However, we observe that for certain values of $f$, the experimental results deviate significantly from the QCRB, with the deviation becoming more pronounced as $f$ increases.
This phenomenon is attributed to the discontinuity of the square wave used in our experiment and the unavoidable phase noise that arises during the detection process.
This explanation is further supported by Monte Carlo simulations~\cite{SupplementalMaterial,Code}.
Notably, this issue can be effectively mitigated when the motion of the point source is continuous.

\begin{figure}[b]
  \centering
  \includegraphics{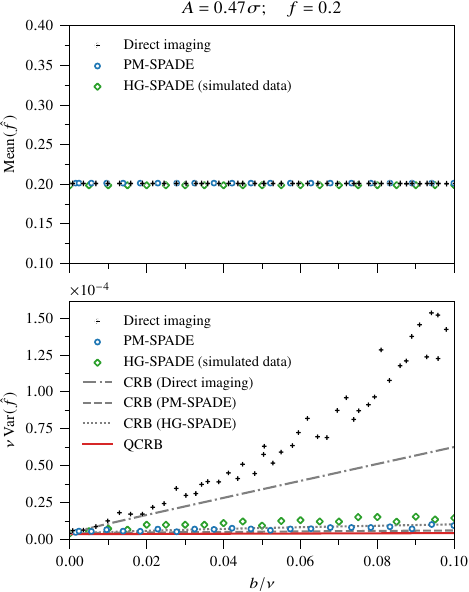}
  \caption{
    Mean and rescaled variance of direct imaging, PM-SPADE, and HG-SPADE versus the excess noise induced by background light.
    The results for HG-SPADE are obtained from simulated data for measuring the first 21 order modes~\cite[Sec.~V]{SupplementalMaterial}.
    Here, the true value of the dimensionless frequency is $f=0.2$ and the oscillation amplitude is $A=0.47\sigma$.
    The Cram\'er-Rao bounds (CRB) are numerically evaluated according to Eq.~\eqref{eq:CFI} and Eq.~\eqref{eq:gamma}, while the QCRB is evaluated according to Eq.~\eqref{eq:noisy_qfi}.
    }
  \label{fig:noisy}
\end{figure}

Figure~\ref{fig:noisy} presents the mean and rescaled variance of frequency estimates obtained through PM-SPADE and direct imaging as functions of the relative intensity of the background noise, which is induced by an LED positioned in front of the camera.
As illustrated in the upper panel of Fig.~\ref{fig:noisy}, the background noise does not compromise the unbiasedness of the frequency estimates. 
The lower panel of Fig.~\ref{fig:noisy} demonstrates that the variance of the frequency estimates obtained by the PM-SPADE measurement is significantly smaller than that from direct imaging, especially under strong background noise conditions.
This finding aligns with the theoretic expectation indicated by the numerical evaluation of the CFI matrices according to Eq.~\eqref{eq:CFI}, which suggests that the PM-SPADE measurement is more robust against excess noise than direct imaging.
In Refs.~\cite{SupplementalMaterial,Code}, we also include the HG-SPADE with the first 21 orders in the comparison via numerical simulations.
We find that HG-SPADE lies between direct imaging and PM-SPADE in terms of the robustness against excess noise at \(A=0.28\sigma\) and \(A=0.47\sigma\). 
This is because, for the HG-SPADE in the sub-Rayleigh regime, the intensity of \(\phi_1\) carries the major information~\cite{Zhou2023} but its signal is small and thus susceptible to noise.
Meanwhile, the PM-SPADE distributes the informative photons more evenly to the two detectors  and thus achieves better noise robustness.

\textit{Conclusion.---}
This study investigated the quantum limit for estimating the motion characteristics of an optical point source in the sub-Rayleigh regime.
We demonstrated that, by elaborately designing the spatial modes to be decomposed, the PM-SPADE outperforms direct imaging as well as the HG-SPADE in estimating the micro-oscillation frequency of an optical point source with excess noise such as dark counts and background lights.
It is worth noting that the advantage of PM-SPADE holds primarily in the sub-Rayleigh regime, as it only measures two specific spatial modes;
For oscillations whose amplitudes are far larger than the characteristic width of the PSF, direct imaging and HG-SPADE are more appropriate for estimating the oscillation frequency.
Our results, on one hand, advance the application of SPADE from static-parameter estimation to dynamic-parameter estimation; on the other hand, they reveal the effectiveness of SPADE in enhancing robustness against noise.
In addition, our work provides a framework for estimating any motion characteristic of an optical point source, so it is also applicable to other parameters such as velocity, acceleration, and even the diffusion coefficient of Brownian motion trajectories, which we will explore in future work.

The ability of PM-SPADE to effectively resist noise and enhance measurement precision makes it a valuable tool for various applications.
Although our implementation of the PM-SPADE uses a CMOS camera, only two pixels are actually employed for photon counting.
This simplicity suggests that, in practice, the CMOS camera could be replaced with just two single-pixel detectors without compromising measurement functionality.
Therefore, the PM-SPADE offers a simple yet effective approach to enhancing noise robustness in optical imaging in the sub-Rayleigh regime. 
It holds promising potential for applications in both biology and astrophysics, particularly in scenarios where background noise is a primary factor limiting measurement precision~\cite{Prusti2016,Moeckl2019}.

\begingroup
\let\addcontentsline\devnull

\textit{Acknowledgments.---}
This work is supported by the Innovation Program for Quantum Science and Technology (Grant No. 2024ZD0301000), the National Natural Science Foundation of China (Grants No.\ 92476118 and No.\ 12275062) and the Natural Science Foundation of Zhejiang Province (LY24A050004, LY23A050003).
We acknowledge Ryo Mizuta Graphics for the free optical components pack for scientific illustration in Blender.

\endgroup

\onecolumngrid
\newpage

\setcounter{equation}{0}
\setcounter{figure}{0}
\setcounter{table}{0}
\setcounter{page}{1}
\makeatletter
\renewcommand{\theequation}{S\arabic{equation}}
\renewcommand{\thefigure}{S\arabic{figure}}
\renewcommand{\thetable}{S\arabic{table}}

% \begin{center}
%   \large\bfseries\parskip\z@skip
%   Supplemental Material for \\
%   ``Measuring the Oscillation Frequency Beyond the Diffraction Limit''
%   \bigskip
% \end{center}

\begin{center}
    \huge\parskip\z@skip
    Supplemental Material
\end{center}

\tableofcontents

\section{Derivation of the Fisher information}

For a weak incoherent optical source, the density operator for the optical fields on the image plane in each short coherence time interval can be well approximated as~\cite{supp_Tsang2016}
\begin{equation}
  \rho_t \approx (1-\epsilon) \op{\mathrm{vac}} + \epsilon \op{\psi_t},
\end{equation}
where \(\epsilon \ll 1\) is the average photon number arriving on the image plane, \(\ket{\mathrm{vac}}\) denotes the vacuum state, and \(\ket{\psi_t}\) is the one-photon state with \(t\) denoting the time instant.
Henceforth, we assume that the average photon number \(\epsilon\) is independent of time \(t\).
For a spatially invariant one-dimensional imaging system, the one-photon state can be expressed as
\begin{equation}
  \ket{\psi_t} = \int \psi(x-s(t, \theta)) \ket{x}\dd{x} ,
\end{equation}
where $\ket{x}$ denotes the eigenstates of photon position at the image plane and \(\psi(x)\) the normalized amplitude point-spread function.
Assume that the motion of the optical source is slow enough so that its position is approximately constant during a single exposure time of the camera---a single sampling process.
Therefore, for a single exposure that consists of \(M\) temporal modes, the density operator for the optical fields can be described as \(\rho_t^{\otimes M}\).

\subsection{Quantum Fisher information under ideal conditions}
Denote by \(\mathcal F[\rho]\) the quantum Fisher information (QFI) matrix of the state \(\rho\) with respect to the vector parameter \(\theta=(\theta_1, \theta_2, \ldots)\).
Due to the additivity of the QFI matrix for the tensor products of quantum states~\cite{supp_Liu2020}, we get
\begin{equation} \label{seq:QFIM1}
  \mathcal F[\rho_t^{\otimes M}]
  = M \mathcal F[\rho_t]
  \approx \nu \mathcal F\qty[\op{\psi_t}],
\end{equation}
where \(\nu \equiv \epsilon M\) is the average photon number detected during a single exposure.
Assuming a Gaussian point-spread function
\begin{equation} \label{seq:psf}
  \psi(x)=(2\pi\sigma^{2})^{-1/4} \exp(-\frac{x^2}{4\sigma^2})
\end{equation}
with \(\sigma\) being the characteristic length, the QFI matrix of pure states can be given by
\begin{align}
  \mathcal F[\op{\psi_t}]_{jk} &= 4\qty(\ip{\pdv{\psi_t}{\theta_j}}{\pdv{\psi_t}{\theta_k}} - \ip{\pdv{\psi_t}{\theta_j}}{\psi_t}\ip{\psi_t}{\pdv{\psi_t}{\theta_k}}) \\
  &= 4\qty(\ip{\pdv{\psi_t}{s}} - \ip{\pdv{\psi_t}{s}}{\psi_t}\ip{\psi_t}{\pdv{\psi_t}{s}}) \pdv{s(t,\theta)}{\theta_j} \pdv{s(t, \theta)}{\theta_k} \\
  &= \frac1{\sigma^2} \pdv{s(t,\theta)}{\theta_j} \pdv{s(t, \theta)}{\theta_k}.\label{seq:QFIM2}
\end{align}

Now, let us consider a set \(T = \qty{n/f_s}_{n=0}^{N-1}\) of sampling time instants, where \(f_s\) is the sampling rate.
The total density operator for all the samples is given by
\begin{equation}
  \rho_\mathrm{total} = \bigotimes_{t\in T} \qty(\rho_t^{\otimes M}).
\end{equation}
According to the additivity of the QFI matrix for the tensor products of quantum states, we get
\begin{equation} \label{seq:QFIM3}
  \mathcal F[\rho_\mathrm{total}] = M \sum_{t\in T}  \mathcal F[\rho_t].
\end{equation}
Combining Eqs.~\eqref{seq:QFIM1}, \eqref{seq:QFIM2}, and \eqref{seq:QFIM3}, we obtain the QFI matrix for the total density operator:
\begin{equation} \label{seq:QFIM}
  \mathcal{F}[\rho_\mathrm{total}]_{jk}
  \approx \frac{\nu}{\sigma^2} \sum_{t\in T} \pdv{s(t,\theta)}{\theta_j} \pdv{s(t, \theta)}{\theta_k}.
\end{equation}

\subsection{Classical Fisher information under excess noise}

The performance of a quantum measurement for parameter estimation can be assessed by the Cram\'er-Rao bound (CRB), which is given by the classical Fisher information (CFI) matrix.
At each sampling, the multi-output photon counting measurement will yield the measurement outcome \(m=(m_1, m_2, \ldots)\), where \(m_j\) denotes the integrated photon number for the \(j\)th detector.
Denote by \(\{E_j\}\) the positive-operator-valued measure (POVM) describing the quantum measurement on the one-photon state.
Under the Poisson limit, each \(m_j\) obeys the Poisson distributions~\cite{supp_Tsang2021, supp_Smith2010, supp_Starr2012}
\begin{equation} \label{seq:poisson}
  p(m) = \prod_j p_j(m_j),\quad p_j(m_j) = \exp(-\Lambda_j) \frac{\Lambda_j^{m_j}}{m_j!},
\end{equation}
where \(\Lambda_j = \nu \ev{E_j}{\psi_t}\) is the average photon number collected on the \(j\)th detector during a single sampling process.
Therefore, the CFI of one sample at time \(t\) is given by
\begin{align} \label{seq:poisson_fisher}
  F(t) = \mathbb E\qty[\qty(\pdv{\ln p(m)}{\theta})^2]
  = \sum_j \mathbb E\qty[\qty(\pdv{\ln p_j(m_j)}{\theta})^2],
\end{align}
where \(\mathbb E\) stands for the expectation taken over the probability distribution \(p(m)\).
For a Poisson distribution given by Eq.~\eqref{seq:poisson}, it follows that
\begin{equation}
  \pdv{\ln p_j(m_j)}{\theta} = \pdv{\Lambda_j}{\theta}\qty(\frac{m_j}{\Lambda_j} - 1).
\end{equation}
Substituting the above expression into Eq.~\eqref{seq:poisson_fisher}, we get
\begin{align}
  F(t) &= \sum_j \qty(\pdv{\Lambda_j}{\theta})^2 \mathbb{E}\qty[\qty(\frac{m_j}{\Lambda_j} - 1)^2]\\
  &=\sum_j \frac{1}{\Lambda_j^2} \qty(\pdv{\Lambda_j}{\theta})^2 \mathbb{E}\qty[\qty(m_j - \Lambda_j)^2]\\
  &=\sum_j \frac{1}{\Lambda_j}\qty(\pdv{\Lambda_j}{\theta})^2, \label{seq:poisson_fisher_result}
\end{align}
where we have used \(\mathbb{E}[(m_j - \Lambda_j)^2] = \Lambda_j\) for Poisson distributions in the last equality.
Meanwhile, the statistical model depends on the parameters \(\theta\) to be estimated only through displacement \(s(t, \theta)\).
Consequently, the CFI can be expressed as
\begin{equation} \label{seq:CFI_t}
  F(t) = \qty[\pdv{s(t,\theta)}{\theta}]^2 \sum_j \frac{1}{\Lambda_j}\qty(\pdv{\Lambda_j}{s})^2.
\end{equation}

Now, we analyze the impact of the background light or dark counts on the CFI.
We assume that photon arrivals from the background light or detector dark counts are rare events, which can be modeled using a Poisson distribution.
Furthermore, we assume that these background light or the dark counts are independent of the signal photons for each photon counters.
Let \(b_j\) denote the average photon number from the background light or the dark counts at the \( j \)th photon counter.
Notably, the sum of two independent Poisson random variables results in another Poisson random variable, with its expectation being the sum of the individual expectations~\cite{supp_Wasserman2010}.
Consequently, Eq.~\eqref{seq:CFI_t} remains valid even when background light or dark counts are present, provided we substitute \(\Lambda_j\) by \(\Lambda_j + b_j\).
Since \(b_j\) is independent of the parameter \(\theta\), the derivative of \(\Lambda_j + b_j\) with respect to \( \theta \) is the same as that of \(\Lambda_j\).
This leads us to
\begin{equation}
  F(t) = \qty[\pdv{s(t,\theta)}{\theta}]^2 \sum_j \frac{1}{\Lambda_j + b_j}\qty(\pdv{\Lambda_j}{s})^2,
\end{equation}
which represents the CFI in the presence of background noise from light or dark counts.
Here, \( \Lambda_j = \nu \mu_j \) with \( \mu_j = \ev{E_j}{\psi_t} \) is the probability of an arrival photon detected by the \( j \)th detector during a single sampling process.
Assuming that \( \nu \) is independent of time \( t \), the CFI for the entire sampling process is given by
\begin{equation}\label{seq:CFI}
  F = \nu \sum_{t\in T} \gamma(t, \theta, b) \qty[\pdv{s(t,\theta)}{\theta}]^2
  \qq{with}
  \gamma(t, \theta, b) \equiv \sum_\ell \frac{1}{\mu_\ell + b_\ell/\nu}\qty(\pdv{\mu_\ell}{s})^2.
\end{equation}
For multiparameter estimation, the above formula for the CFI can be straightforwardly generalized to the following CFI matrix
\begin{equation}\label{seq:CFIM}
  F_{jk}
  = \nu \sum_{t \in T} \gamma(t,\theta, b) \pdv{s(t,\theta)}{\theta_j}\pdv{s(t,\theta)}{\theta_k},
\end{equation}
which is Eq.~\eqref{eq:CFI} in the main text.

\subsection{Quantum Fisher information under excess noise}
To derive the quantum limit of the CFI matrix in Eq.~\eqref{seq:CFI}, we can construct an intensity operator~\cite{supp_Tsang2021} \(\Gamma_t\) which satisfies
\begin{equation}
    \tr(\Gamma_t E_j) = \nu\ev{E_j}{\psi_t} + b_j.
\end{equation}
We consider the noise that is caused by detector noise or background light and assume that the noise intensity is independent of the quantum measurements and is uniformly distributed across all measurement outcomes, i.e., \(b_j=b\).
Therefore, we model the quantum state of the excess noise as a maximally mixed state on the Hilbert space \(\mathcal H\) associated with \(d\) spatial modes.
We further assume that the measurement is a one-dimensional orthogonal projective measurement.
For such cases, the intensity operator can be expressed as
\begin{equation}
    \Gamma_t = (\nu + d b) \eta_t
    \qq{with}
    \eta_t = \tau_\mathrm{signal} \op{\psi_t} + \tau_\mathrm{noise} \mathds{1},
\end{equation}
where \(\tau_\mathrm{signal} = \nu / (\nu + d b)\), \(\tau_\mathrm{noise} = b / (\nu + d b)\), and \(\mathds{1}\) is the \(d\)-dimensional identity operator.
Then, the CFI in Eq.~\eqref{seq:CFI} is bounded by the QFI \(\mathcal F\) given by
\begin{equation}
    \mathcal F = (\nu + d b) \sum_{t \in T} \mathcal F[\eta_t],
\end{equation}
where \(\mathcal F[\eta_t]\) is the QFI of the one-photon state \(\eta_t\).
Assuming that \(\eta_t = \sum_j \lambda_j \op{e_j}\) is the eigenvalue decomposition, the QFI of \(\eta_t\) about \(\theta\) is given by~\cite{supp_Liu2020}
\begin{equation}
    \mathcal F[\eta_t] = \qty(\pdv{s}{\theta})^2 \sum_{jk} \frac{2}{\lambda_j+\lambda_k} \qty|\mel{e_j}{\pdv{\eta_t}{s}}{e_k}|^2.
\end{equation}
In our model, we may choose the first two eigenvectors of \(\eta_t\) as follows:
\begin{equation}
    \ket{e_1} = \ket{\psi}\qand\ket{e_2}=\frac{\ket{\psi'}}{\sqrt{\ip{\psi'}}}=2\sigma\ket{\psi'},
\end{equation}
whose corresponding eigenvalues are given by \(\lambda_1 = \tau_\mathrm{signal} + \tau_\mathrm{noise}\) and \(\lambda_j=\tau_\mathrm{noise}\) for \(j>1\).
Note that
\begin{equation}
    \pdv{\eta_t}{s} = \tau_\mathrm{signal} \pdv{\op{\psi_t}}{s}
    = \frac{\tau_\mathrm{signal}}{2\sigma} (\op{e_2}{e_1} + \op{e_1}{e_2})
\end{equation}
and
\begin{equation}
    \mel{e_j}{\pdv{\eta_t}{s}}{e_k} = 0 \qif j > 2 \qor k > 2.
\end{equation}
It then follows that
\begin{align}
    \mathcal F[\eta_t] &=  \qty(\pdv{s}{\theta})^2 \frac{4}{\lambda_1 + \lambda_2} \qty|\mel{e_1}{\pdv{\eta_t}{s}}{e_2}|^2 \\
    &=\qty(\pdv{s}{\theta})^2 \frac{\tau_\mathrm{signal}^2}{\sigma^2(\tau_\mathrm{signal} + 2 \tau_\mathrm{noise})}\\
    &=\qty(\pdv{s}{\theta})^2 \frac{\nu^2}{\sigma^2(\nu + db)(\nu+2b)}.
\end{align}
Therefore, we obtain the QFI for the noisy case as
\begin{equation}
    \mathcal F = \frac{\nu}{\sigma^2} \frac{1}{1+2b/\nu} \sum_{t\in T} \qty(\pdv{s(t,\theta)}{\theta})^2.
\end{equation}
The above QFI can be easily generalized the QFI matrix
\begin{equation}\label{seq:QFIM_noisy}
    \mathcal F_{jk} = \frac{\nu}{\sigma^2} \frac{1}{1+2b/\nu} \sum_{t\in T}
    \pdv{s(t,\theta)}{\theta_j} \pdv{s(t,\theta)}{\theta_k}
\end{equation}
for the multiparameter estimation case.

\section{Practical measurement strategies}

We here consider three different measurement strategies: direct imaging (DI), the HG-SPADE, and the PM-SPADE.
Let $\ket{\phi_q}$ denote the one-photon state in the \(q\)th-order Hermite-Gaussian (HG) mode, whose wave function is given by
\begin{equation}\label{seq:HG_q}
  \phi_q(x) = \qty(\frac1{2\pi\sigma^2})^{1/4}\frac1{\sqrt{2^qq!}}H_q\qty(\frac x{\sqrt2\sigma})\exp\qty(-\frac{x^2}{4\sigma^2}),
\end{equation}
where $H_q$ is \(q\)th-order the Hermite polynomial and $\sigma$ is the characteristic width of the point-spread function.
The PM modes are defined as the following superpositions of the two lowest-order HG modes:
\begin{equation}\label{seq:PM_pm}
  \ket{\phi_\pm} \equiv \frac1{\sqrt 2} (\ket{\phi_0} \pm \ket{\phi_1}).
\end{equation}
The corresponding POVM elements for these three measurements are given by
\begin{equation}
  E_k^\mathrm{(DI)} = \int_{ka-a/2}^{ka+a/2}\op{x}\dd{x},\quad
  E_q^\mathrm{(HG)} = \op{\phi_q},\quad
  E_\pm^\mathrm{(PM)} = \op{\phi_\pm},
\end{equation}
where $k \in \mathbb{Z}$, $q \in \mathbb{N}$, and \(a\) is the pixel size of direct imaging.
The POVM elements for the SPADE with the PM modes are incomplete, but we still calculate the CFI matrix by Eq.~\eqref{seq:CFIM}; This means that we discard the information about all other modes that are orthogonal to the PM modes.

For the Gaussian point-spread function given by Eq.~\eqref{seq:psf}, we have
\begin{align}
  \mu^\mathrm{(DI)}_k
  &=\frac12 \operatorname{erfc}\qty(\frac{s - k a - a / 2  }{\sqrt2\sigma})
  - \frac12 \operatorname{erfc}\qty(\frac{s - k a + a / 2}{\sqrt2\sigma}),\nonumber \\
  \mu^\mathrm{(HG)}_q
  &= \frac{1}{q!} \qty(\frac{s}{2\sigma})^{2q} \exp(-\frac{s^2}{4\sigma^2}),\nonumber \\
  \mu^\mathrm{(PM)}_\pm
  &= \frac12\qty(\frac{s}{2\sigma} \pm 1)^2\exp(-\frac{s^2}{4\sigma^2}), \label{seq:mu}
\end{align}
where $\operatorname{erfc}$ is the complementary error function.
The CFI matrix for the three measurements can be straightforwardly calculated by Eq.~\eqref{seq:CFIM} with the above expressions, at least in a numerical way.

When the excess noise is absent, i.e., $b_j = 0$ for all \(j\), we have
\begin{align} \label{seq:gamma}
  \lim_{a\to0} \gamma^\mathrm{(DI)} = \gamma^{\mathrm{(HG)}} = \frac1{\sigma^2},
  \qand
  \gamma^\mathrm{(PM)} = \frac1{\sigma^2}\qty[ 1 - \qty(\frac{s}{2\sigma})^2 + \qty(\frac{s}{2\sigma})^4 ] \exp(-\frac{s^2}{4\sigma^2}),
\end{align}
where \( \lim_{a \to 0} \) means the limit when the pixel size \(a\) approaches infinitesimal.
Notice that \(\gamma^\mathrm{(PM)}\) is close to \( 1/\sigma^2 \) when the range of \(s\) is very small, indicating that the SPADE with the PM modes are near-optimal for in the ideal case when the displacements of the point source are small.
Since we are primarily concerned with objects moving in the Rayleigh limit, this condition is naturally satisfied in our experiment.
It can be seen from Eq.~\eqref{seq:gamma} that \( \lim_{a\to0} \gamma^\mathrm{(DI)} = \gamma^{\mathrm{(HG)}} \geq \gamma^\mathrm{(PM)} \) holds for each single sampling, which in turn implies that
\begin{equation} \label{seq:CFI_relation}
  \mathbf v^\mathsf{T} \mathcal{F} \mathbf v
  = \mathbf v^\mathsf{T} F^\mathrm{(DI)} \mathbf v
  = \mathbf v^\mathsf{T} F^\mathrm{(HG)} \mathbf v
  \geq \mathbf v^\mathsf{T} F^\mathrm{(PM)} \mathbf v
\end{equation}
with \( \mathbf v \) being an arbitrary column vector of the same dimension as the parameter vector \( \theta \).
Since Eq.~\eqref{seq:CFI_relation} holds for all \( \mathbf v \), we can conclude that
\begin{equation}
  \mathcal{F} = F^\mathrm{(DI)} = F^\mathrm{(HG)} \geq F^\mathrm{(PM)}
\end{equation}
for the ideal case without excess noise.

\begin{figure}[!ht]
  \centering
  \includegraphics{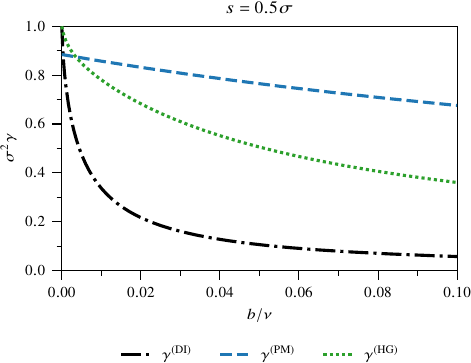}
  \caption{Values of $\gamma$ for direct imaging, HG-SPADE, and PM-SPADE.
  Here, the pixel size for direct imaging is set to be \(a=4.6\)\,{\textmu}m.}
  \label{sfig:gamma}
\end{figure}

Now, we consider the case where the background noise is present and uniform for all detectors.
We numerically calculate the factor \(\gamma\) for the three measurements and present the results in Fig.~\ref{sfig:gamma}.
The larger the \(\gamma\) factor, the better the measurement scheme is robust against the excess noise.
It is demonstrated from Fig.~\ref{sfig:gamma} that the PM-SPADE measurement has the highest robustness against excess noise.

Moreover, for the limiting case of infinitesimal-amplitude motion (i.e., \(s(t,\theta) \ll \sigma\) for all \(t\)), the PM-SPADE is optimal for estimating the motion characteristic \(\theta\) in the presence of excess noise.
For the PM-SPADE under the limiting case of \(s\to 0\), we have
\begin{equation}
    \lim_{s\to0} \mu_+^{\mathrm{(PM)}} = \lim_{s\to0} \mu_-^{\mathrm{(PM)}} = 1/2 \qand
    \lim_{s\to0} \sum_{\ell=\pm} \frac1{\mu_\ell^{\mathrm{(PM)}}} \qty(\pdv{\mu_\ell^{\mathrm{(PM)}}}{s})^2 = \frac1{\sigma^2}.
\end{equation}
Therefore, we obtain
\begin{align}
    \lim_{s\to0}\gamma^{\mathrm{(PM)}}(t, \theta, b)
    &=  \frac{1}{1+2b/\nu} \lim_{s\to0} \sum_\ell \frac{1}{\mu_\ell}\qty(\pdv{\mu_\ell}{s})^2
    = \frac{1}{\sigma^2}\frac{1}{1+2b/\nu},
\end{align}
implying that \(\lim_{s\to0}{F^\mathrm{(PM)}} = \mathcal{F}\) under the noise being considered in this work.

\section{Estimation of oscillation frequency}

Assume that the point-like optical source undergoes a simple harmonic motion:
\begin{equation}
  s(t,\theta) = A \sin(2\pi f_o t + \varphi) + A,
\end{equation}
where we have chosen the minimum position of the oscillation as the origin of the coordinate system.
Let \(T = \qty{n/f_s}_{n=0}^{N-1}\) be a set of sampling time instants with \( f_s \) being the sampling rate.
We take \(f\equiv f_o/f_s\) as the parameter of interest and assume that the values of \(A\) and \(\varphi\) are known.
For such a single parameter estimation problem, according to Eq.~\eqref{seq:QFIM}, the QFI about \(f\) can be expressed as
\begin{align}
  \mathcal F &= \frac{4\pi^2 \nu A^2}{\sigma^2} \sum_{n=0}^{N-1} n^2 \cos^2(2\pi f n + \varphi) \nonumber \\
  &= \frac{4\pi^2 \nu A^2}{\sigma^2} \sum_{n=0}^{N-1} \frac{n^2}{2} \qty(1 + \cos(4\pi f n + 2\varphi)).
\end{align}
Assuming that $f$ is not near 0 or $1/2$, we can use the following approximation (see Ref.~\cite{supp_Stoica1989} or Ref.~\cite[Sec.~3.11]{supp_Kay1993}):
\begin{align} \label{seq:approx}
  \frac1{N^3}\sum_{n=0}^{N-1} n^2\cos(4\pi f n + 2\varphi) \approx 0
\end{align}
for a large \(N\).
Using the above approximation and the formula \(\sum_{n=0}^{N-1} n^2 = N (N-1) (2 N-1) / 6\) for the sum of squares, the QFI can be expressed as
\begin{equation}
  \mathcal F \approx \frac{\nu \pi^2 A^2}{3\sigma^2} N (N-1) (2 N-1).
\end{equation}
Correspondingly, we get the quantum Cram\'er-Rao bound (QCRB) for estimating the frequency \(f\):
\begin{equation} \label{seq:QCRB_sin}
  \operatorname{Var}(\hat f) \gtrsim \frac{3\sigma^2}{\nu \pi^2 A^2 N(N-1)(2N-1)}.
\end{equation}
For the ideal case without background noise, direct imaging with infinitesimal pixel size and the HG-SPADE measurement can achieve the QCRB, while the PM-SPADE measurement can only approach the QCRB in the limit of small motion regions.

Note that for the cases where the background noise is present and uniform for all detectors with infinitesimal pixel size, \( \gamma^\mathrm{(DI)} \) is independent of the value of \( s \);
This is due to the uniformity of the background noise and the shift invariance of the direct imaging measurement.
Therefore, the CFI for direct imaging can be expressed as
\begin{align}
  F^\mathrm{(DI)} &= \nu \gamma^\mathrm{(DI)} \sum_{n=0}^{N-1} \qty[\pdv{s(n/f_s,\theta)}{\theta}]^2
  \approx \frac{\nu \pi^2 A^2\gamma^\mathrm{(DI)}}{3} N (N-1) (2 N-1),
\end{align}
where we have used the same approximation Eq.~\eqref{seq:approx} as before.

\begin{figure}[ht]
  \centering
  \includegraphics{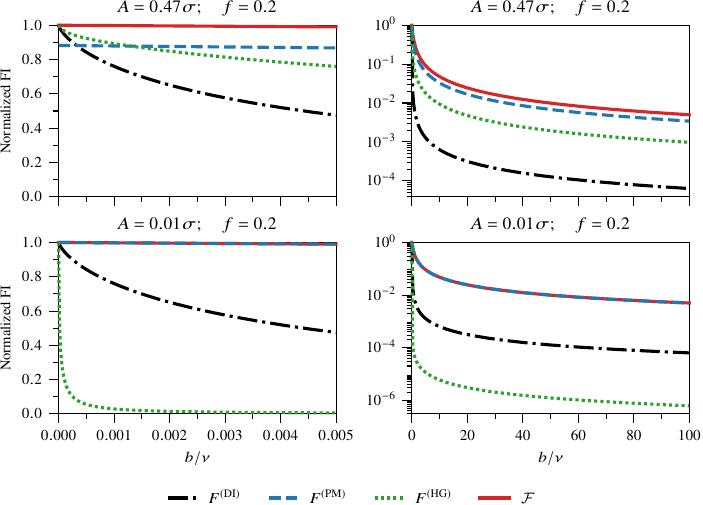}
  \caption{
      The various FI about the dimensionless oscillation frequency $f$ versus the relative noise intensity $b/\nu$.
      The FI is calculated using Eq.~\eqref{seq:CFIM}, Eq.~\eqref{seq:mu}, and Eq.~\eqref{seq:QFIM_noisy}, and normalized with respect to that under ideal conditions without excess noise.
      The upper row of Fig.~\ref{sfig:fi} corresponds to the parameter settings used in our experiment, while the lower row shows results obtained with a smaller oscillation amplitude, specifically $ A = 0.01\sigma $.
      The left column corresponds to the noise levels used in our experiment, whereas the right column illustrates cases with extremely high excess noise.
  }
  \label{sfig:fi}
\end{figure}

Figure~\ref{sfig:fi} shows the Fisher information (FI) for the oscillation frequency under different oscillation amplitudes and excess noise levels.

It is difficult to simulate a sinusoidal micro-oscillation of the optical point source using the DMD, as the DMD is a digital device with a discrete set of pixels.
In our experiment, we use the square wave function
\begin{equation}\label{seq:square_wave}
  s(t,\theta) = A\operatorname{sgn} \qty[\sin(2\pi f_o t)] + A
\end{equation}
to mimic the sinusoidal oscillation.
Using the Fourier expansion, we can rewrite the square wave function as
\begin{align}
  s(t,\theta) &= \frac{4A}{\pi} + \frac{4A}{\pi} \sum_{k=1}^{\infty} \frac{\sin[2\pi (2k-1) f_o t]}{2k-1} \\
  &= \frac{4A}{\pi} + \frac{4A}{\pi} \qty[\sin(2\pi f_o t) + \frac{1}{3}\sin(6\pi f_o t) + \frac{1}{5}\sin(10\pi f_o t) + \ldots]. \label{seq:square_wave_fourier}
\end{align}
For the data sampled at time instant \(t\), we first estimate the displacement \(s(t)\) by the maximum likelihood estimation  method and denote by \(\hat{s}(t)\) the estimates.
We then estimate the frequency \(f\) by the least squares estimation  method with \(\{ \hat{s}(t) \mid t \in T \}\) as the observed values and the first term of the Fourier expansion, \( (4 A / \pi) \sin(2\pi f_o t) \), as the model function.
In such case, the QCRB is given by Eq.~\eqref{seq:QCRB_sin} with \( A \) being replaced by \( 4A/\pi \), that is,
\begin{equation}
  \operatorname{Var}(\hat f) \gtrsim \frac{3\sigma^2}{ 16 \nu A^2 N(N-1)(2N-1)}.
\end{equation}
This is Eq.~\eqref{eq:QCRB} in the main text.

\section{Experimental details}

The experimental setup is shown in Fig.~\ref{fig:setup} of the main text.
We here provide additional details on the whole experimental process, including the simulation of point source motion, the spatial light modulation, and the camera readout.
The key instruments used in the experiment, with their main specifications and the settings employed during the experiment, are listed below:
\begin{itemize}
    \item \textbf{Laser:} Photodigm, 770DBRL-T08, continuous-wave emission at a wavelength of 770\,nm.
    \item \textbf{DMD:} VIALUX, V-7001 VIS, $1024\times768$ micromirrors, 13.7\,{\textmu}m pitch, maximum frame rate of 22,727\,Hz (44\,{\textmu}s of picture time).
    \item \textbf{SLM:} HoloEye, PLUTO-2-NIR-011, $1920\times1080$ pixels, 8\,{\textmu}m pitch, input frame rate 60\,Hz, wavelength range of 420--1100\,nm, phase-only modulation.
    \item \textbf{CMOS camera:} Hamamatsu, ORCA-Quest qCMOS, C15550-20UP, $4096\times2304$ pixels, 4.6\,{\textmu}m pitch, quantum efficiency 55.28\,\% at 770\,nm, sampling rate $f_s = 20$\,Hz, exposure time 0.5\,ms.
\end{itemize}

\subsection{Simulation of point source motion }

We simulate the point source by illuminating a continuous wave laser with a wavelength of \(\lambda = 770\,\mathrm{nm}\) (Photodigm, 770DBRL-T08) onto a DMD (VIALUX, V-7001 VIS with 1024 $\times$ 768 micromirrors of 13.7\,{\textmu}m size each) with only one micromirror flipped at a time.
The square-wave oscillation of a point source, i.e.,
\begin{equation}
    s(t, \theta) = A \operatorname{sgn}[\sin(2\pi f_o t)] + A,
\end{equation}
where $A$ and $f_{o}$ represent the amplitude and the frequency of the oscillation, respectively, is rendered by flipping different micromirrors according to the given pattern.
We use the above square-wave oscillation to mimic the sinusoidal oscillation, which is the first order term of the Fourier expansion Eq.~\eqref{seq:square_wave_fourier}.

Our task is to infer the motion characteristics of the light source from the light field obtained from a diffraction-limit imaging system.
We use an iris and Lens\,1 to form a unit-magnification diffraction-limited imaging system,
whose point-spread function is approximated by a Gaussian function with a characteristic width of $\sigma \approx 103$\,{\textmu}m.
The light emitted from the single point source is directed sequentially through the iris with a diameter of 0.8\,mm and Lens\,1 (focal length $l_{1}=200$\,mm) positioned at a distance of $2l_{1}$ from the DMD.

\subsection{Spatial light modulation}

We implemented the PM-SPADE measurement using a digital holographic technique~\cite{supp_Paur2016,supp_Zhou2023}.
A phase-only SLM (HoloEye, PLUTO-2-NIR-011 with 1920 $\times$ 1080 pixels of pitch 8\,{\textmu}m) was positioned in the image plane of the imaging system and modulates the phase of incoming optical fields by adjusting the voltage at each pixel to rotate the liquid crystal molecules, thereby changing the pixel's effective refractive index~\cite{supp_RosalesGuzman2017}.
In other words, a phase-only SLM transforms the incoming optical field \(U(x,y)\) as
\begin{equation}
    U(x,y) \to U'(x,y) = U(x,y)\exp[i\mathcal{G}(x,y)],
\end{equation}
in terms of the computer-generated hologram (CGH) \(\mathcal G(x,y)\) loaded onto the SLM.

A phase-only SLM can indirectly realize a phase modulation on each pixel, which together with a Fourier lens can be used for complex amplitude modulation and mode decomposition of the incoming optical field.
Suppose that we intend to realize a complex amplitude modulation whose polar form is
\begin{equation} \label{seq:V_polar_form}
    V(x,y)= a(x,y) e^{i\varphi(x,y)},
\end{equation}
where \(a\) takes values in the interval \([0,1]\) and \(\varphi\) takes values in \([-\pi, \pi]\).
At each pixel \((x,y)\), the CGH can be treated as a function of the amplitude \(a\) and the phase \(\varphi\).
Since \(\mathcal G\) is \(2\pi\)-periodic with respect to \(\varphi\), it can be expanded as a Fourier series in the form
\begin{equation} \label{seq:Fourier_series}
    e^{i\mathcal{G}} = \sum_{m=-\infty}^{\infty} c_m e^{im\varphi},
\end{equation}
where only the first-order term will be utilized for complex amplitude modulation and separated by the technology of spatial carrier frequency~\cite{supp_RosalesGuzman2017}.
In our work, we use the following algorithm for the CGH~\cite{supp_Arrizon2007}:
\begin{equation}
    \mathcal{G}=f(a) \sin(\varphi),
\end{equation}
where $f$ is a nonlinear mapping to be defined later.
It follows from the Jacobi-Anger identity that
\begin{equation}\label{seq:U_prime}
    e^{i \mathcal G} = e^{if(a)\sin\varphi}
    = \sum_{m=-\infty}^{\infty} \mathcal{J}_m[f(a)]e^{im\varphi},
\end{equation}
where $\mathcal{J}_m$ is the $m$th order Bessel function of the first kind.
The function \(f\) is chosen such that $\mathcal{J}_1[f(a)]=\kappa a$, where the constant factor $\kappa \approx 0.5819$, the maximal value of \(\mathcal{J}_1\), restricts \(a\) to be not larger than \(1\).
As a result, the first-order term of Eq.~\eqref{seq:U_prime} becomes $\kappa V(x,y)$.
In our work, to decompose an optical field onto the transverse spatial modes $\phi_+(x,y)$ and $\phi_-(x,y)$, we use the following modulation function:
\begin{equation} \label{seq:V_concrete}
    V(x, y) = \frac1{V_\mathrm{max}} \qty[\phi_+^*(x, y) e^{i k_y y} + \phi_-^*(x, y) e^{- i k_y y}] e^{i k_x x},
\end{equation}
where $k_x$ and $k_y$ are the spatial carrier frequencies and \(V_\mathrm{max}\) is the factor that makes the maximum value of \(|V(x,y)|\) be unity.
Comparing Eq.~\eqref{seq:V_concrete} with Eq.~\eqref{seq:V_polar_form}, we have
\begin{equation} \label{seq:phi}
    \varphi(x,y) = \varphi'(x,y) + k_x x
    \qq{with}
    \varphi'(x,y) \equiv \arg[\phi_+^*(x, y) e^{i k_y y} + \phi_-^*(x, y) e^{- i k_y y}],
\end{equation}
where \(\arg\) denotes the argument of complex numbers.
The spatial frequency $k_x$ is used to isolate the first-order term of Eq.~\eqref{seq:U_prime} in the horizontal direction at the Fourier plane, while $k_y$ is used to isolate the component modulated by \(\phi_+\) and \(\phi_-\) in the vertical direction at the Fourier plane.

\subsection{Camera readout}

At the Fourier plane of a lens (Lens\,2 in Fig.~\ref{fig:setup} in the main text) with a focal length of $l_2=150$\,mm,  after neglecting a global phase and the coordinate-independent factor, the optical field is given by~\cite{supp_Goodman2003}
\begin{equation}
    \widetilde U(x',y') \propto \mathscr{F}[U'(x,y)]\qty(\omega_x,\omega_y)
    \qq{with}
    \omega_x \equiv \frac{2\pi x'}{\lambda l_2} \qand \omega_y \equiv \frac{2\pi y'}{\lambda l_2},
\end{equation}
where \(\mathscr{F}[U(x,y)]\qty(\omega_x,\omega_y)\) is the two-dimensional Fourier transform give by
\begin{equation}
    \mathscr{F}[U'(x,y)]\qty(\omega_x,\omega_y)
    \equiv \frac1{2\pi} \iint U'(x,y) \exp[- i(\omega_x x + \omega_y y)] \dd{x} \dd{y}.
\end{equation}
With the modulation given by Eq.~\eqref{seq:Fourier_series} and Eq.~\eqref{seq:phi}, we have
\begin{align}
    \mathscr{F}[U'(x,y)]\qty(\omega_x,\omega_y)
    &= \sum_{m=-\infty}^\infty \mathscr{F}\qty[U(x,y) c_m(x,y) e^{im\varphi'(x,y) + i m k_x x}]\qty(\omega_x,\omega_y) \\
    &= \sum_{m=-\infty}^\infty \mathscr{F}\qty[U(x,y) c_m(x,y) e^{im\varphi'(x,y)}]\qty(\omega_x - m k_x, \omega_y),
\end{align}
meaning that the spatial carrier frequency \(k_x\) shifts the \(m\)th-order components by \(m k_x \lambda l_2 / (2\pi)\) at the Fourier plane.
Assume that \(k_x\) is sufficiently large such that the first-order term can be spatially separated from all other terms.
Therefore, at the Fourier plane where \(\omega_x = k_x\), we have
\begin{align}
    \mathscr{F}[U'(x,y)]\qty(\omega_x=k_x,\omega_y)
    &\propto  \mathscr{F}\qty[U(x,y)\phi_+^*(x,y) e^{ i k_y y}](0,\omega_y)
    + \mathscr{F}\qty[U(x,y)\phi_-^*(x,y) e^{-i k_y y}](0,\omega_y) \\
    &= \mathscr{F}\qty[U(x,y)\phi_+^*(x,y)](0,\omega_y - k_y)
    + \mathscr{F}\qty[U(x,y)\phi_-^*(x,y)](0,\omega_y + k_y).
\end{align}
When \(k_y\) is sufficiently large, the two terms above can be spatially separated in the vertical direction.
Let us define \(x_0'\equiv\lambda l_2 k_x / (2\pi)\) and \(y'_0\equiv\pm \lambda l_2 k_y / (2\pi)\).
At the point \((x'_0,\pm y'_0)\) in the Fourier plane, the optical intensities satisfy
\begin{align}
    I(x'_0, y'_0) &\propto \abs{\mathscr{F}\qty[U(x,y)\phi_+^*(x,y)](0,0)}^2
    = \abs{\iint \phi_+^*(x,y) U(x,y) \dd{x}\dd{y}}^2,  \\
    I(x'_0, -y'_0) &\propto \abs{\mathscr{F}\qty[U(x,y)\phi_-^*(x,y)](0,0)}^2
    = \abs{\iint \phi_-^*(x,y) U(x,y) \dd{x}\dd{y}}^2.
\end{align}
Therefore, measuring the intensity at \((x_0, y_0)\) and \((x_0,-y_0)\) in the Fourier plane can give us the proportion between  the photon occupancies of the PM modes.

To extract information about the photon occupancy of the PM modes, only the first order diffraction light from the SLM was directed to a CMOS camera (Hamamatsu, ORCA-Quest qCMOS, C15550-20UP, 4096 $\times$ 2304 pixels with a 4.6\,{\textmu}m pitch, and a quantum efficiency of 55.28\,\% at 770\,nm) with a sampling rate of $f_s=20$\,Hz, which was placed at the Fourier plane of Lens\,2.
The camera enables real-time monitoring of photon occupancy in the PM modes by reading out the photon count at two specific pixels.
The exposure time of the CMOS camera was set to 0.5\,ms, resulting in approximately 60 photons recorded per frame by the PM-SPADE.
We utilized 50 frames, sampled over a total duration of 2.5\,s, to perform a frequency estimation.
For each frame, we used the maximum likelihood estimator (MLE) to estimate the displacement of the point source.
Subsequently, the least squares estimation (LSE) is used to infer the oscillation frequency of the point source.

\section{Monte Carlo simulation}\label{sec:mc_sim}

\subsection{Explanation for the abnormal increase of estimates' variance}

\begin{figure}[b]
  \centering
  \includegraphics{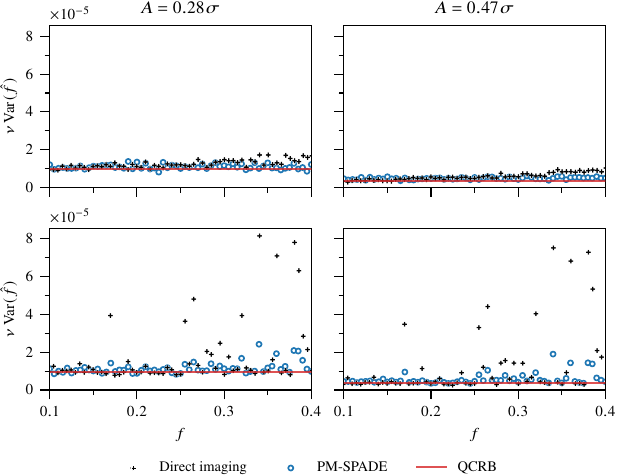}
  \caption{
    Simulation results for different types of motions with random delays in the initial frame's sampling timing.
    In the top two panels, the optical point source undergoes sinusoidal motion, while in the bottom two panels, it undergoes square wave motion.
    We here set the random delay $\delta \tau$ to a normal distribution with a mean of 2.8\,ms and a standard deviation of 0.48\,ms.
  }
  \label{sfig:simulation}
\end{figure}

Figure~\ref{fig:ideal} in the main text shows that the experimental variance of frequency estimates for direct imaging exhibits an unexpected  rise when compared to the QCRB at specific frequency points.
We identify two primary causes for this discrepancy: (i) random delays in the initial frame's sampling timing, and (ii) the discontinuous nature of the square wave function used to simulate the sinusoidal oscillation in our experiment.
We shall explain these two factors in what follows.

In our setup, the software-triggered signal sent to the CMOS camera was susceptible to timing variations due to program execution delays and USB port latency.
This resulted in a random delay in the sampling moment of the initial frame, which was not accounted for in our theoretical analysis.
In this work, we use the square wave function to simulate the sinusoidal oscillation of the optical point source, due to the limitations of the DMD.
The random delay at the sampling moment of the initial frame can be modeled by a random variable $\delta\tau$, with which the time-dependent displacement of the optical point source is given by
\begin{equation} \label{app_SquareWave}
  s^{\rm(sgn)}(t,\theta) = A\operatorname{sgn}\qty[\sin(2\pi f_o (t + \delta\tau))] + A.
\end{equation}
Since the square wave function is discontinuous when the optical point source is at the transition points, the random delay $\delta\tau$ can cause a significant change in the estimated displacement and thus in the estimated frequency.

To support our explanation, we performed a simulation to investigate the impact of the random delay on the variance of the frequency estimates.
The code of the simulations can be found in our GitHub repository~\cite{supp_Code}.
In our simulation, we set the random delay $\delta \tau$ to a normal distribution.
The detector pixel size is set to 4.6\,{\textmu}m, matching the specifications of the CMOS camera used in the experiment.
We configured the signal photon number to 400 for the direct imaging measurement and 60 for the PM-SPADE measurement, closely aligning with the photon counts observed in the experimental data.
Considering that the square wave motion Eq.~\eqref{app_SquareWave} is used to simulate the sinusoidal motion corresponding to the first terms in the Fourier expansion, we also perform the simulation for the time-dependent displacement
\begin{equation}
  s^{\rm(sin)}(t,\theta) = \frac{4A}{\pi} \sin(2\pi f_o (t + \delta\tau)) + \frac{4A}{\pi}.
\end{equation}
As shown in the top two panels of Fig.~\ref{sfig:simulation}, there is no obvious abnormal increase in the variance of the frequency estimates for the genuine sinusoidal motion.
For the square wave motion used in this work to simulate the sinusoidal motion, the bottom two panels of Fig.~\ref{sfig:simulation} show that the abnormal increase in the variance of the frequency estimates can be produced by introducing the random delay of the initial frame's sampling timing.

\subsection{Hermite-Gaussian-mode SPADE}

We here include the HG-mode measurement in the comparison.
Since implementing HG-SPADE experimentally is challenging, we use Monte Carlo simulations to support our theoretical results.
The simulation code is available in our GitHub repository~\cite{supp_Code}.

\begin{figure}[!ht]
\centering
\includegraphics{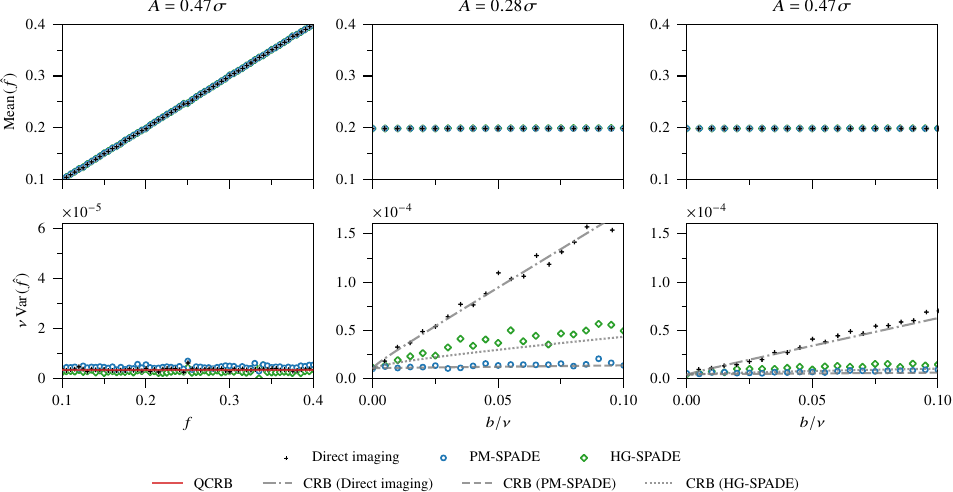}
\caption{
    Simulation results for different measurement schemes.
    The left two panels plot the mean and variance of frequency estimates for each measurement scheme without excess noise, while the others plot the results under excess noise conditions.
    The vertical axes are scaled to match those in the main text for better visibility.
}
\label{sfig:hgspade}
\end{figure}

We simulated the photon occupancy of first 21 Hermite-Gaussian modes and applied the maximum likelihood estimator for the position of optical point sources.
By combining the estimated positions over time, we reconstructed a time-domain signal, from which we estimate the oscillation frequency using LSE (same procedure as direct imaging and PM-SPADE).
We plot the results in Fig.~\ref{sfig:hgspade}, which shows that the theoretical results agree well with our simulations.
For the cases without excess noise as shown in the left two panels of Fig~\ref{sfig:hgspade}, the HG-SPADE and direct imaging are both optimal for estimating the oscillation frequency, while PM-SPADE is slightly suboptimal.
The robustness of HG-SPADE against the excess noise lies between that of direct imaging and PM-SPADE.
In addition, the advantage of PM-SPADE is most evident when the oscillation amplitude is small (e.g., \(A=0.28\sigma\)), because its estimation performance deteriorates for large displacements.

\section{Intuitive understanding on the noise robustness of PM-SPADE}

\begin{figure}[!ht]
\centering
\includegraphics{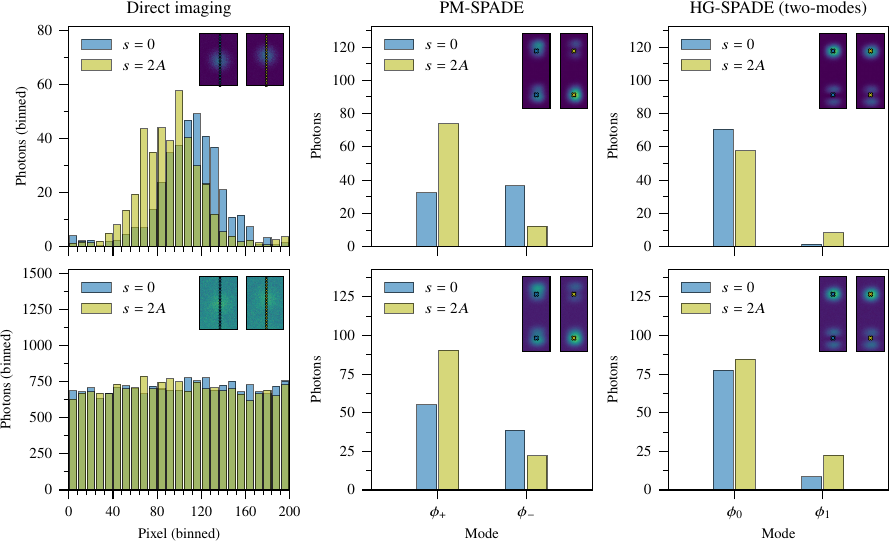}
\caption{
    Data collected from the CMOS camera for different positions of the optical point source.
    The top three panels show the camera data when no intentional noise is introduced, while the bottom three panels plot the camera data with the excess noise level $b/\nu \approx 0.1$.
    In each panel, the inset displays the image of the CMOS camera for \(s=0\) (left) and \(s=2A\) (right).
    For direct imaging, only the data along the dotted line are used, whereas for the PM-SPADE and the HG-SPADE, only the data marked with crosses are used.
    The histograms for direct imaging have been binned in groups of 8 pixels for better visibility.
}
\label{sfig:intuitive}
\end{figure}

The superior noise robustness of the PM-SPADE can be intuitively understood by examining the data collected from the CMOS camera, as shown in Fig.~\ref{sfig:intuitive}.
The precision of oscillation-frequency estimation is strongly relevant to the precision of estimating the light source position at each time instant.
Therefore, we focus on the distinction between the maximum and minimum positions of the optical point source reflected in the collected data.
In each panel of Fig.~\ref{sfig:intuitive}, the inset displays the image of the CMOS camera for \(s=0\) (left) and \(s=2A\) (right).
Direct imaging distinguishes the position of the optical point source via the shift of intensity distribution along the dotted line, see the top-left panel; 
With excess noise, these two images become difficult to distinguish, as shown in the bottom-left panel.
The PM-SPADE and HG-SPADE, on the other hand, manifest the distinction via the intensities of some specific pixels (marked with crosses).
The PM-SPADE maintains a clear distinction between \(s=0\) and \(s=2A\)  even under excess noise conditions, as shown in the bottom-middle panel.
For comparison, we also perform the HG-SPADE with the first two Hermite-Gaussian modes.
The HG-SPADE manifests the major distinction via the intensity of the \(\phi_1\) mode for \(s\) in the sub-Rayleigh regime~\cite{supp_Zhou2023}.
Nevertheless, the signal of the \(\phi_1\) mode intensity is small and thus more susceptible to noise than that of the PM-SPADE.
Meanwhile, the PM-SPADE distributes the informative photons more evenly to the two detectors, so it achieves better noise robustness than the HG-SPADE.

\end{document}